\documentclass[prd,aps,amsmath,amssymb,letter,showpacs]{revtex4}
\usepackage{graphicx}% Include figure files
\usepackage{epsfig,rotate,latexsym}
\usepackage{hyperref}
\begin{document}
\title{Domain growth and fluctuations during quenched transition to 
QGP in relativistic heavy-ion collisions}
\author{Ranjita K. Mohapatra}
\email {ranjita@iopb.res.in}
\author{Ajit M. Srivastava}
\email{ajit@iopb.res.in}
\affiliation{Institute of Physics, Sachivalaya Marg, 
Bhubaneswer 751005, India}

\begin{abstract}
We model the initial confinement-deconfinement transition in relativistic
heavy-ion collisions as a  rapid quench in view of expected rapid
thermalization to a QGP state. The transition is studied using
the Polyakov loop model, with the initial field configuration (in the 
confining phase) covering a small neighborhood of the confining vacuum 
$l \simeq 0$, as appropriate for $T < T_c$. Quench is implemented by
evolving this initial configuration with the effective potential at
a temperature $T > T_c$.
We study the formation of Z(3) domain structure and its 
evolution during the transition as $l$ rolls down in different directions 
from the top of the central hill in the effective potential of $l$.
When explicit Z(3) symmetry breaking effects (arising from dynamical 
quark effects) are small, then we find well defined Z(3) domains which
coarsen in time. Remarkably, the magnitude plot of $l$ shows vacuum bubble 
like configurations arising during the quench. This first order transition
like behavior occurs even though there is no metastable vacuum separated 
by a barrier from the true vacuum for the parameter values used.
When the initial field configuration everywhere rolls down roughly along 
the same direction (as will happen with large explicit symmetry breaking) 
then we do not find such bubble-like configurations. However, in this case 
we find huge oscillations of $l$ with large length scales. We show that  
such large oscillations can lead to large fluctuations in the evolution 
of flow anisotropies compared to the equilibrium transition case.
\end{abstract}

\maketitle

\section{INTRODUCTION}

 In relativistic heavy-ion collision experiments (RHICE) the collision of 
two nuclei leads to a hot dense region which is expected to rapidly achieve
state of thermal equilibrium. For the relevant range of energies and
colliding nuclei at RHIC and at LHC there is compelling evidence that
a region of quark-gluon plasma is created in these collisions. Simulation
results as well as experimental data, such as elliptic flow measurements,
all point towards a very rapid thermalization to the QGP phase, within
a proper time less than 1 fm. We thus have a system which starts out in the
confining phase, and within proper time of (probably much less than)  
1 fm makes a transition to the QGP phase (with maximum temperature
estimates ranging from 200 MeV to more than 700 MeV for the relevant energies
in these experiments). Lattice results show that for real world QCD 
with small baryon density (as appropriate for the central rapidity 
regions in RHICE) the transition is likely to be a crossover. With  that,
the dynamics of the transition depends crucially on the rate of
temperature change compared to the time scale of the evolution of
the order parameter field. For an equilibrium transition, we had studied
the formation of Z(3) domains and associated {\it QGP strings}
using a first order transition dynamics via bubble nucleation 
\cite{gupta,gupta2}. The transition was simulated using the Polyakov loop, 
$l(x)$, as an order parameter for the confinement-deconfinement 
transition \cite{plkv}. These studies were appropriate for large chemical
potential case, as in lower energy collisions, where the transition is 
expected to be of first order. Though, the results for Z(3) wall network 
etc. having certain universal characters,  may be applicable
in more general context, even for a cross-over, as explained 
in \cite{gupta,gupta2}.
  
  However, given the very short time scale of initial thermalization
to QGP state, an equilibrium dynamics of the transition appears
unlikely. A more appropriate description of the transition should employ
quenched dynamics in which the growth of Z(3) domains will be via
spinodal decomposition. In this paper we carry out such a study 
using the Polyakov loop, $l(x)$, as an an order parameter for the 
confinement-deconfinement transition, with an effective potential of
the kind used in refs.\cite{plkvall,psrsk,psrsk2}. For our simulation 
results we choose a definite parametrization of the effective 
potential \cite{psrsk} as was used in our previous works \cite{gupta,gupta2}.
We model the phase transition in this Polyakov loop model, with 
the initial field configuration (in the confining phase) covering a small 
neighborhood of the confining vacuum $l \simeq 0$ as appropriate for
the initial $T$ = 0 system. In a quench, the temperature rapidly (rather 
suddenly)  increases to its maximum value $T_0$ with the effective
potential changed accordingly.  The initial field $l$, unable to relax
to the new equilibrium vacuum state in this short time, becomes
unstable and rolls down in different directions from the top of the 
central hill in the effective potential of $l$. We study the formation 
of Z(3) domain structure during this evolution. 
When explicit Z(3) symmetry breaking effects (arising from dynamical 
quark effects) are small, then we find well defined Z(3) domains which
coarsen in time. With a symmetric initial patch of $l$, all the three
Z(3) domains form with random shapes and rapidly increase in size by
coarsening.  Remarkably, the magnitude plot of $l$ shows vacuum bubble 
like configurations, such as those which arise in a first order transition,
arising during the quench in this case (when the initial field rolls down 
in different directions). This first order transition like behavior 
occurs even though there is no metastable vacuum separated 
by a barrier from the true vacuum for the parameter values used. 
These bubble like configurations expand as well, somewhat in similar
manner as during a first order transition.  

When the initial patch of $l$ is only partially symmetric around $l = 0$ 
(as appropriate for small explicit symmetry breaking from quark effects), 
the dynamics retains these qualitative aspects with expected changes. 
Thus, true vacuum domains (with $\theta = 0$) are more abundant and
they also grow fast at the cost of the other two Z(3) domains (which
are now metastable due to explicit symmetry breaking). Still, for
small explicit symmetry breaking, all the three types of domains occur
during the initial stages of the transition. There are few, small
Z(3) walls as there are fewer metastable Z(3) domains embedded in
the dominant true vacuum with $\theta = 0$. These walls shrink rapidly
and eventually only the true vacuum survives.

The dynamics is found to be very different when the explicit symmetry
breaking due to quark effects is taken to be strong. In this case the initial 
patch of $l$ (around equilibrium point for $T = 0$ effective potential) 
could be significantly shifted towards the true vacuum for the quenched
$T = T_0$ effective potential. In such a situation, $l$ will 
roll down roughly along the same direction with angular variations becoming
smaller during the roll down. In this case only $\theta = 0$ vacuum
survives and no other Z(3) domains are formed. Also, in this case we 
do not find bubble-like configurations. However, in this case we find 
huge oscillations of $l$ with large length scales. This behavior is known
from previous studies of the dynamics of scalar field in a quench \cite{infl}
and is expected here when angular variations are small. The dynamics
of field in such a case is dominated by large length scale coherent
oscillations. It leads to novel scenarios of reheating via parametric
resonance in the case of inflation in the early Universe \cite{iinfl}. In 
our case of RHICE also it raises important questions about the possibility 
of parametric resonance and of novel
modes of particle production from these large oscillations of $l$ during
the early stages of the transition. In the present work we explore another
important effect of these large oscillations, on the evolution of flow
anisotropies in RHICE. As most of the flow anisotropies are expected to
develop during first few fm of the QGP formation, it becomes an important
question if the presence of large oscillations of $l$ can affect the
development of these flow anisotropies. As we will see, this indeed happens
and these large oscillations lead to large fluctuations in flow anisotropy.

We mention here that we do not discuss which of the cases discussed 
here (small, or large explicit symmetry breaking) may actually be
realized in RHICE. This is primarily because of lack of understanding of
the magnitude of explicit symmetry breaking term near $T = T_c$.
(Some discussion of this has been provided in our earlier works
\cite{gupta,gupta2}.) Also, the spread of initial field configuration
about confining vacuum plays a crucial role here and that in turn
depends on details of pre-equilibration stage. A proper understanding
(or modeling) of this stage and resulting estimate of the initial
spread is essential before one can make more definitive statements
about the dynamics of the transition. We hope to discuss these issues
in future.  Our main purpose here is to illustrate the possibility
of very different types of dynamics of transition depending on the
initial configuration. Our results show that in quenched dynamics, Z(3) 
domains, and resulting Z(3) domain walls, can last for any reasonably
length of time only when explicit symmetry breaking terms are very
small. Otherwise either different domains do not form at all, hence
no Z(3) walls are formed, or fewer metastable Z(3) domains form embedded 
in most abundant true vacuum region. In the latter case resulting
Z(3) walls are smaller to begin with, and disappear quickly.

  Importance of Z(3) walls in RHICE has been discussed by us in
previous works, where we have also emphasized non-trivial scattering of
quarks from Z(3) walls. We have explored its consequences for 
cosmology as well as for RHICE \cite{layek,gupta,gupta2,ananta}, including
the possibility of CP violating scattering of quarks from Z(3) walls
leading to interesting observational implications \cite{atreya}. 
Recently an interesting possibility has been discussed by Asakawa 
et al. in ref. \cite{bass} where it is argued that scattering of 
partons from Z(3) walls may account for small viscosity as well as
large opacity of QGP. Our results for the formation of Z(3) domains
during quench (which are abundant only for small explicit symmetry 
breaking case) can be relevant for the studies in ref. \cite{bass}. In 
this context we mention that smallest reasonable size Z(3) domains (hence 
Z(3) walls) we find are of order 1-2 fm at very early times,  and at 
that stage the magnitude
of the Polyakov loop order parameter $l$ is very small, of order
few percent of its vacuum expectation value. The quark scattering from 
Z(3) walls is likely to be small for such a small magnitude of $l$
\cite{layek,ananta,atreya}. By the time the magnitude of $l$ becomes
significant, domains coarsen to have large sizes, of order several fm.
Thus, in the context of our model it appears difficult to form
very small Z(3) domains which still can scatter partons effectively
(as needed in the study of ref.\cite{bass}).

 It is important to note that our results for domain growth and fluctuations
for the quench case are dominated by the spinodal instabilities during
the roll down of the field from the top of the hill of the effective
potential. These instabilities arise primarily from the nature of
the quench when the initial field configuration becomes unstable
to exponential growth of long wavelength modes due to sudden change in
the shape of the effective potential (the quench), and will be in general 
present even if the transition is a cross-over. In this sense we believe 
that the qualitative aspects of our results have a wider applicability and 
are not crucially dependent on the specific form of the effective 
potential used here.  

 We briefly mention here that issues related to the physical meaning of
Z(3) domains etc. have been discussed in the literature (see, discussion
of these in our earlier works \cite{gupta,gupta2} and we will not repeat
it here. We only note that recent work of Deka et al. \cite{dglltc} 
has provided a support for the existence of these metastable vacua from 
Lattice.  We also note that a simulation of spinodal decomposition in 
Polyakov loop model has been carried out in ref. \cite{dumitru}, where 
fluctuations in the Polyakov loop and growth of long wavelength modes
(representing domain formation) are investigated. In comparison, 
the main focus of our work is on detailed growth of domains due to 
coarsening with and without explicit symmetry breaking, new bubble like 
structures, and existence of large fluctuations  affecting flow 
anisotropies in important ways.

       The paper is organized in the following manner. In section II, 
we discuss the essential aspects of the Polyakov loop model of 
confinement-deconfinement phase transition and the effective potential 
used from ref.\cite{psrsk}. Section III presents the numerical technique 
of simulating the phase transition via quench. We first discuss the case 
without any explicit symmetry breaking and study the formation and
evolution of Z(3) domains during the quench. Sec. IV discusses the
unexpected result of bubble like structures arising during the quench.
We also discuss the case of small explicit symmetry breaking term when
true vacuum domains become more abundant, but overall picture of
the transition remains roughly similar.
In Sec.V we discuss the case when explicit symmetry breaking effects
are strong leading to $l$ rolling down everywhere roughly  along 
$\theta = 0$. This leads to large oscillations of $l$. In Sec. VI
we study the effects of these large oscillations on flow anisotropy 
and show that it leads to large fluctuations in elliptic flow and
spatial eccentricity (as compared to the case of
equilibrium transition). Section VIII presents conclusions.

\section{MODELING THE PHASE TRANSITION}

 We briefly recall the salient features of the model we use for
the confinement-deconfinement phase transition. The order parameter 
is taken to be the the expectation value of Polyakov loop $l(x)$,

\begin{equation}
  l(\vec{x}) ={1 \over N} Tr({\cal P} exp(ig\int^\beta_0 
A_0(\vec{x},\tau)d\tau))
\end{equation}

Where $A_0(\vec{x},\tau)$ is the time component of the vector potential 
$A_{\mu}(\vec{x},\tau)=A^{a}_{\mu}(\vec{x},\tau) T^a$, $T^a$ are the 
generators of $SU(N)$ in the fundamental representation,  $\cal{P}$ denotes 
path ordering in the Euclidean time $\tau$, g is the gauge coupling, and
$\beta = 1/T$ with T being the temperature. $N$ (= 3 for QCD) is the 
number of colors. The complex scalar field $l(\vec{x})$ 
transform under the global $Z(N)$ (center) symmetry transformation as 

\begin{equation}
 l(\vec{x}) \rightarrow {exp(2{\pi}in/N) l(\vec{x})}, \: n = 0,1,...(N-1)
\label{eq:zns}
\end{equation}

 The expectation value of $l(x)$ is related to $e^{-\beta F}$ where
$F$ is the free energy of an infinitely heavy test quark.
For temperatures below $T_c$, in the confined phase, the expectation value
of Polyakov loop  is zero corresponding to the infinite
free energy of an isolated test quark. (Hereafter, we will use the
same notation $l(x)$ to denote the expectation value of the Polyakov
loop.)  Hence the Z(N) symmetry is restored below $T_c$. Z(N) symmetry
is broken spontaneously above $T_c$ where $l(x)$ is non-zero
corresponding to the finite free energy of the test quark. Effective
theory of the Polyakov loop has been proposed by several authors
with various parameters fitted to reproduce lattice results for
pure QCD \cite{plkvall,psrsk,psrsk2}. We use the Polyakov loop effective 
theory proposed by Pisarski \cite{psrsk, psrsk2}. The effective 
Lagrangian density can be written as

\begin{equation}
L={N\over g^2} |{\partial_\mu l}|^2{T^2}- V(l)
\end{equation}

Where the effective potential $V(\it l)$ for the Polyakov loop, in case 
of pure gauge theory is given as

\begin{equation}
 V(l)=({-b_2\over2} |l|^2- {b_3\over 6}( l^3+( l^ \ast)^
 3)+\frac{1}{4}(|l|^2)^2){b_4{T^4}}
\end{equation}

At low temperature where $\it{l} = 0$, the potential has only one 
minimum. As temperature becomes higher than $T_c$ the Polyakov loop 
develops a non vanishing  vacuum expectation value $l_0$, and 
the $cos3\theta$ term, coming from the $l^3 + l^{*3}$ term above 
leads to $Z(3)$ generate vacua. Now in the deconfined phase, for a small
range of temperature above $T_c$, the $\it{l} = 0$ extremum becomes the local 
minimum (false vacuum) and a potential barrier exist between the local 
minimum and global minimum (true vacuum) of the potential.
(However, the quench temperature used here is much higher than this
range and there is no barrier present at such temperatures.)

The effects of dynamical quarks is included in terms of explicit breaking 
of the $Z(3)$ symmetry which is represented in the effective potential by 
inclusion of a linear term in $l$ \cite{psrsk, psrsk2, z3lnr}. The 
potential of Eq.(4) with the linear term becomes,

\begin{equation}
 V(l)=\Bigl( -\frac{b_1}{2}(l + l^*) -\frac{b_2}{2} |l|^2- 
{b_3\over 6} (l^3+ {l^{\ast}}^3) +\frac{1}{4}(|l|^2)^2 \Bigr){b_4{T^4}}
\end{equation}

Here coefficient $b_1$ measures the strength of explicit symmetry breaking. 
The coefficients $b_1$, $b_2$, $b_3$ and $b_4$ are dimensionless 
quantities. With $b_1 = 0$, other parameters $b_2$, $b_3$ and $b_4$ are
fitted in ref.\cite{psrsk,psrsk2} such that that the effective potential 
reproduces the thermodynamics of pure $SU(3)$ gauge theory on lattice 
\cite{z3lnr,scav,latt}. The values of various coefficients from refs.
\cite{psrsk,psrsk2} are the same as used in our previous works 
\cite{gupta,gupta2} (including discussions about explicit symmetry breaking
strength $b_1$) and we do not repeat that discussion here. (With those
values of parameters, the transition temperature is taken to be
$T_c = $ 182 MeV.)

\section{NUMERICAL SIMULATION}

In this work, we carry out a $2+1$ dimensional field theoretic simulation 
of the dynamics of confinement-deconfinement transition in a quench.
First we work  within the framework of Bjorken's boost invariant 
longitudinal expansion model \cite{bjorken} (without any transverse
expansion) for the central rapidity region in RHICE. To model 
the quench, we take the initial field configuration
to constitute a small patch around $l = 0$ which corresponds to
the vacuum configuration for the confining phase. This is for the case
of zero explicit symmetry breaking. We will discuss the case of explicit
symmetry breaking later. We have taken the
initial phase of $l$ to vary randomly between 0 and $2\pi$ from one
lattice site to the other, while the magnitude of $l$ is taken to
vary uniformly between 0 and $\epsilon$. (We have also taken the initial
magnitude to have a fixed value equal to $\epsilon$ and the results are 
similar.) Value of $\epsilon$ is taken to be much smaller than the vacuum 
expectation value (vev) of $l$ at $T = T_0$ and results remain 
qualitatively the same for similar small values of $\epsilon$. We report 
results for $\epsilon = 0.01$ times the vev of $l$. We take the 
quench temperature  $T_0 = 400$ MeV. 

  This initial field configuration, which represents the equilibrium
field configuration of a system with $T < T_c$, is evolved using
the effective potential with $T = T_0 > T_c$. This represents the
transition dynamics  of a quench.  The field configuration is evolved
by time dependent equation of motion in the Minkowski space \cite{rndrp}

\begin{equation}
  \frac{\partial^{2}l_j}{\partial\tau^{2}} + \frac{1}{\tau} 
\frac{\partial l_j}{\partial \tau}  -\frac{\partial ^{2}l_j}{\partial x^{2}}   
-\frac{\partial^{2}l_j}{\partial y^{2}} 
= -\frac{g^2}{2NT^2} \frac{\partial{V(l)}}{\partial{l_j}} ;
\quad j = 1, 2
\label{eq:evolution}
\end{equation}

with $\frac{\partial l_j}{\partial \tau} = 0$ at $\tau = 0$ and $l = 
l_1 +i l_2$. The temperature is taken to decrease starting with
the value $T_0$ (at an initial time $\tau_0$ which we take to
be 1 fm in all the simulations) as $\tau^{-1/3}$ as appropriate for the 
longitudinal expansion model. We mention that later in Sect.V we will
consider an isotropic geometry for the transverse dynamics of QGP
as relevant for RHICE. There we will also model non-zero transverse 
expansion and then the central temperature will decrease faster than 
$\tau^{-1/3}$.  Here, we take a $2000 \times 2000$ square lattice 
with physical size 20 fm x 20 fm. We take this lattice as representing 
the transverse plane of the QGP formed in a central collision and consider 
mid rapidity region. The evolution of field was numerically implemented 
by a stabilized leapfrog algorithm of second order accuracy both in space 
and in time with the second order derivatives of $l_i$ approximated by a 
diamond shaped grid \cite{gupta,gupta2}. We evolve the field using 
the periodic, fixed, and free boundary conditions for the square lattice. 
Here we present the results with periodic boundary conditions. We use $\Delta 
x = 0.01$ fm for the present case, later on we use different values of 
$\Delta x$, as shown by the lattice size in the corresponding figures.
we take $\Delta t = \Delta x/\sqrt 2$  as well as $\Delta t = 
0.9 \Delta x/\sqrt 2$ to satisfy the Courant stability criteria. The stability 
and accuracy of the simulation is checked using the conservation of 
energy during simulation. The total energy fluctuations remains few percent 
without any net increase or decrease of total energy in the absence of 
dissipative $\dot l$ term in Eq.(6).

  First, we present results for the formation and evolution of Z(3) domains.
The initial field configuration in the neighborhood of $l = 0$ becomes 
unstable when evolved with the effective potential (Eq.(5)) with $T = T_0
= 400$ MeV (at $\tau = \tau_0 = $ 1 fm). As $l$ rolls down in different 
direction, settling in one
of the three Z(3) vacua, different Z(3) domains form. Initially, as the
phase of $l$ is taken to vary randomly from one lattice site to the next,
there are no well defined domains.  Also, the magnitude of $l$ is very small
initially making any association of Z(3) structure meaningless at such
early stages. Situation remains similar for very early times as seen in 
Fig.1a at an early stage $\tau = 1.2$ fm (i.e. 0.2 fm after the quench). 
In Fig.1 we have shown the
values of the phase of $l$ around the three Z(3) vacua in terms of different
shades (colors) to focus on the evolution of Z(3) domain structure. Thus
all the values of the phase $\theta$ of $l$ are separated in three ranges,
between $-2\pi/6$ to $2\pi/6$ ($\theta = 0$ vacuum), between $2\pi/6$ to 
$\pi$ ($\theta = 2\pi/3$ vacuum), and between $\pi$ to $2\pi - 2\pi/6$
($\theta = 4\pi/3$ vacuum).  As the field magnitude 
grows, the angular variation of $l$ also becomes less random over small 
length scales, leading to a sort of Z(3) domain structure. Z(3) domains
become more well defined, and grow in size by coarsening as shown in
sequence of figures Fig.1b-d. Different shades (colors) in Fig.1 represent 
the three Z(3) domains. Fig.1b-d  shows the growth of these domains at 
$\tau$ = 2.0, 2.4, and 2.8 fm,  as $l$ relaxes to the three Z(3) vacua and 
domains grow in size by coarsening. The magnitude of $l$ is about 0.04, 
0.08, 0.2, and 0.4 for Figs.1a,b,c, and d respectively. Note that domains 
grow rapidly to size of order 2 fm within a time duration of about 1 fm as 
shown in Fig.1b. Within another 1 fm time, domain size is about 4 fm as 
seen in Fig.1d.

The boundaries of different Z(3) domains
represent Z(3) walls, and junction of three different Z(3) domains 
gives rise to the QGP strings. These objects have been discussed in
detail in our earlier works \cite{layek2,gupta,gupta2}.   

\begin{figure*}[!hpt]
\begin{center}
\leavevmode
\epsfysize=8truecm \vbox{\epsfbox{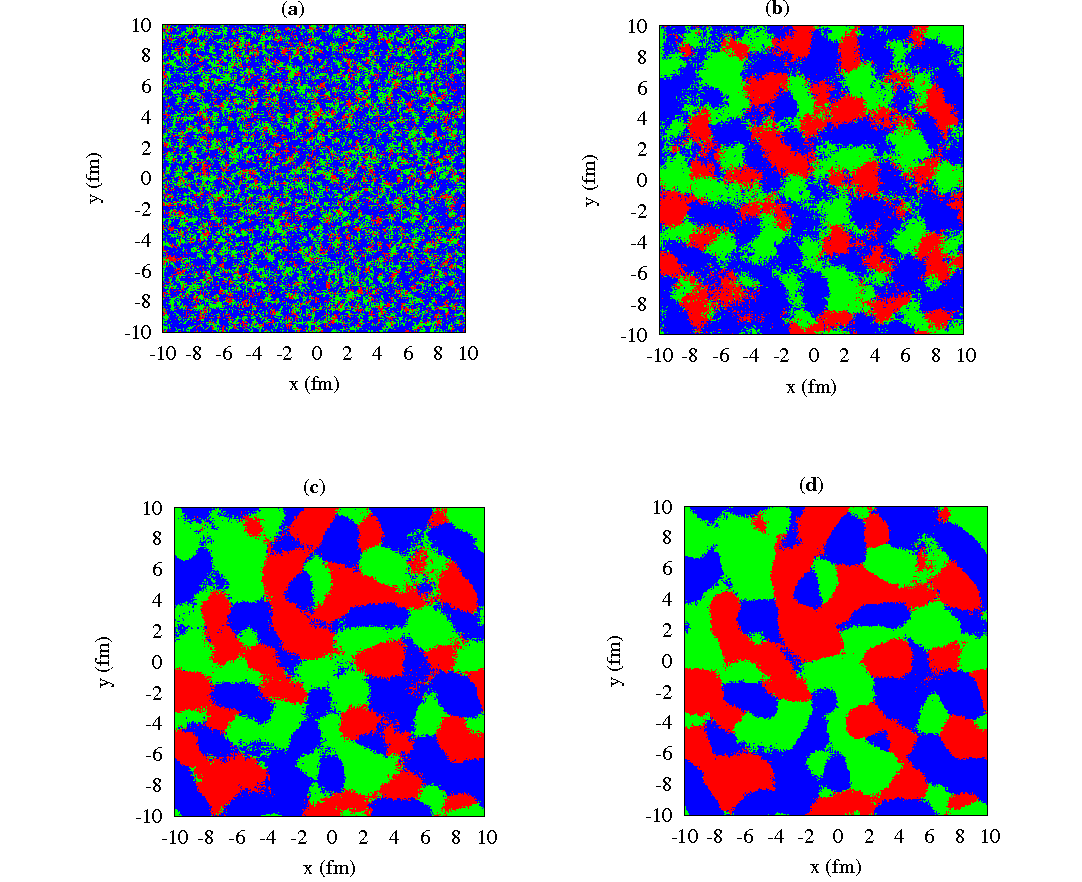}}
\end{center}
\caption{}{(a) Evolution of Z(3) domain structure during a quench.
Note we take the initial temperature to be $T_0$ (= 400 MeV) at proper time
$\tau = \tau_0 = 1$ fm to set the initial conditions for the simulation. 
Different shades (colors) represent the three Z(3) domains. (a) - (d) 
show the growth of these domains at $\tau$ = 1.2, 2.0, 2.4, and 2.8 fm,  
(with corresponding values of temperature $T$ = 376, 317, 298, 283 MeV)
as $l$ relaxes to the three Z(3) vacua and domains grow
in size by coarsening. The magnitude of $l$ is about 0.04, 0.08, 0.2,
and 0.4 for (a), (b), (c), and (d) respectively. Note that domains grow 
rapidly to size of order 2 fm within a time duration of about 1 fm as 
shown in (b). Within another 1 fm time, domain size is about 4 fm (as 
seen in (d)).}
\label{Fig.1}
\end{figure*}

We mention here that smallest reasonable size Z(3) domains (hence Z(3) 
walls), which we find in our simulation, are of order 1-2 fm at very early 
times, such as seen in Fig.1b. At this stage, the magnitude of the Polyakov 
loop order parameter $l$ is still very small, of order few percent of its 
vacuum expectation value. This is important when one considers the 
possibility of nontrivial scattering of partons from Z(3) walls 
\cite{layek,ananta,atreya,bass}. The quark scattering from 
Z(3) walls is likely to be small for such a small magnitude of $l$
\cite{layek,ananta,atreya}. By the time the magnitude of $l$ becomes
significant, domains coarsen to have large sizes, of order several fm,
as in Fig.c,d where the magnitude of $l$ is about 20\% and 40\%, 
respectively, of its vacuum
expectation value. Thus, in the context of our model, with quenched
dynamics of transition, it appears difficult to form
very small Z(3) domains which still can scatter partons effectively
(as needed in the study of ref.\cite{bass}).

\section{BUBBLE LIKE STRUCTURES DURING QUENCH}

 In this section we show a very unexpected result. Fig.2 shows sequence of
plots of the magnitude of $l$ during the quench described in the
previous section. We note the appearance of bubble like structures.
These structures are also seen to grow in a manner similar to the bubbles
for a conventional first order phase transition, as in ref.\cite{gupta,gupta2}.
However, the distribution of the phase of $l$ does not show any specific
local variation related to these bubble-like configurations. In a roughly 
uniform region of the phase these localized bubble-like configurations
arise and expand. We show here the plots of $l$ in Fig.2  corresponding
to the initial central temperature $T = T_0 = $ 500 MeV (at
$\tau = \tau_0 = $ 1 fm). The plots in Fig.2a,b,c,d  are for 
$\tau =$ 2.0, 2.8, 3.0, and 3.3 fm respectively, with the corresponding
values of the central temperature being $T = $ 395, 355, 345, 336 MeV.
It is important to note that at such high temperatures there is no 
metastable confining vacuum at $l = 0$ in the effective potential 
\cite{gupta2}. (Metastable confining vacuum exists from $T = T_c = 182$
MeV upto $T \simeq 250$ MeV.)  There is no tunneling modeled, (or 
thermal hopping over the barrier) here, nor is expected. One expects a 
simple roll down of the field representing the dynamics of spinodal 
decomposition during the quench. We mention that similar bubble-like
configurations also arise with $T_0 = 400$ MeV. However, in that case
the temperature range goes below $T = 250$ MeV. To make a clear
case that these bubble like configuration have nothing to do with
a first order transition like situation (even remotely), we have
shown plots with $T_0 = $ 500 MeV.

\begin{figure*}[!hpt]
\begin{center}
\leavevmode
\epsfysize=8truecm \vbox{\epsfbox{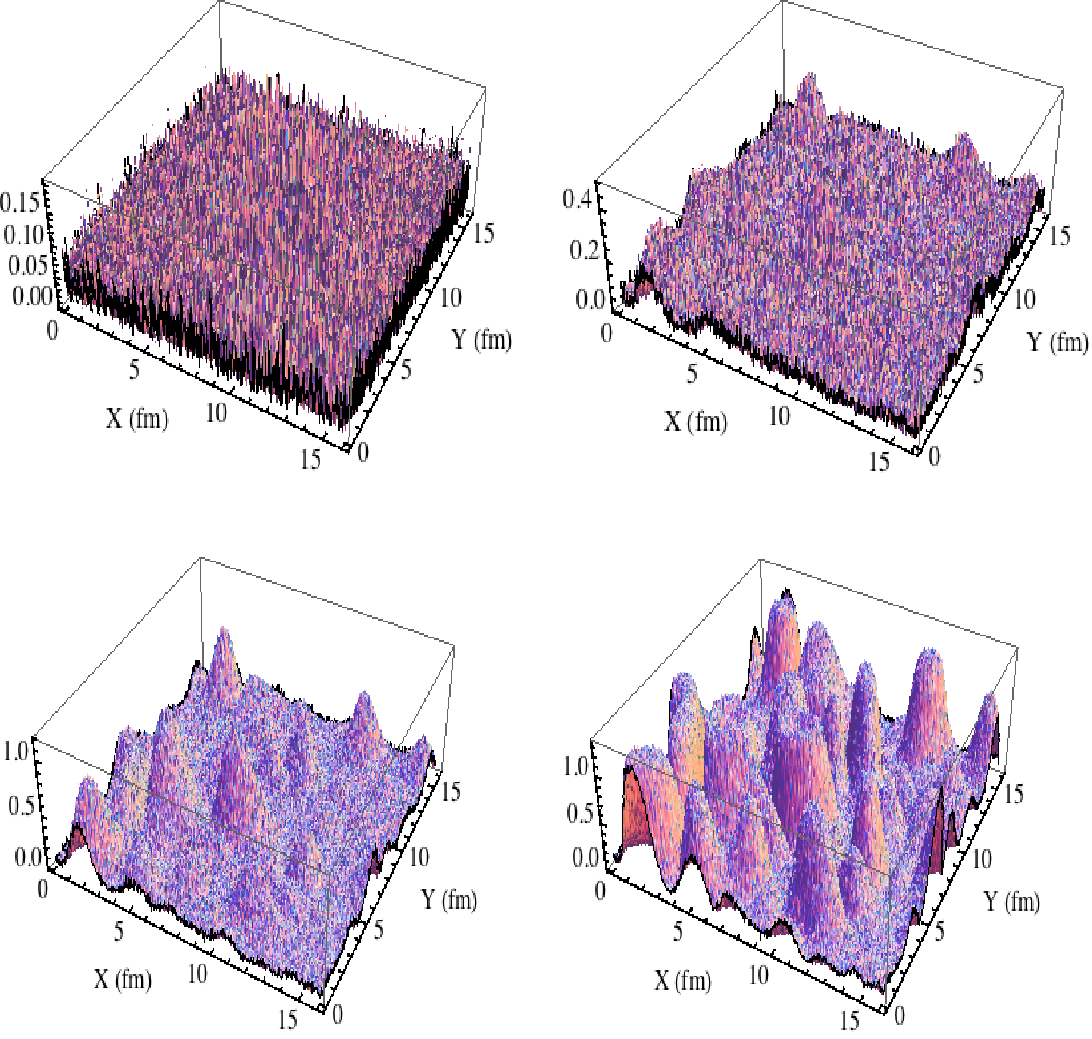}}
\end{center}
\caption{}{These plots correspond to initial central temperature $T_0 = $
500 MeV. (a)-(d) show the surface plots of $|l|$ during early stages
with subsequent  formation and growth of bubble like structures
just like a first order phase transition. Plots in (a) -(d)  are 
at $\tau = $ 2, 2.78, 3 and 3.3 fm respectively, with corresponding
temperature being $T = $ 395, 355, 345, 336 MeV. These bubble-like
configurations are surprising as there is no barrier here, and no 
metastable vacuum in the effective potential for this temperature range.}
\label{Fig.2}
\end{figure*}

This result is very unexpected and points to new interesting 
possibilities for the phase transition dynamics. For example, such bubble
like structures may lead to a dynamics of phase separation in the
case of the universe similar to the original Witten's scenario 
\cite{witten}, even when there is no underlying first order transition.
This may also have important implications for RHICE. More importantly
this new possibility of transition dynamics needs to be understood
and analyzed in detail, see \cite{bubble} for a study of these issues.  

 The studies of this section and the previous section apply to
the case without any explicit symmetry breaking, i.e. with $b_1 = 0$
in Eq.(5). We have repeated these simulations with small explicit
symmetry breaking effects on the initial conditions. We take $b_1 = 
0.005$ as in ref.\cite{gupta2} here as well as in the next section.
By small explicit symmetry breaking effects on the initial conditions
we mean that the initial patch of $l$ is taken to shift towards
$\theta =0$ vacuum for $T = T_0$ effective potential, while
still overlapping with the initial equilibrium value of $l$. This
forces $l$ to roll down to different $\theta$ directions at least
at some fraction of lattice points, though major fraction now
rolls down towards $\theta = 0$. Fig.3 shows sequence of plots showing
growth of Z(3) domains in one such case. We see that one of the vacua
($\theta = 0$) expands dominantly while other domains remain 
relatively smaller. The Z(3) domain walls in this case are smaller 
and disappear faster compared to the case without explicit symmetry
breaking. The dynamics of bubble like structure retains its 
qualitative aspects in this case as long as the field rolls down in 
different directions.

\begin{figure*}[!hpt]
\begin{center}
\leavevmode
\epsfysize=8truecm \vbox{\epsfbox{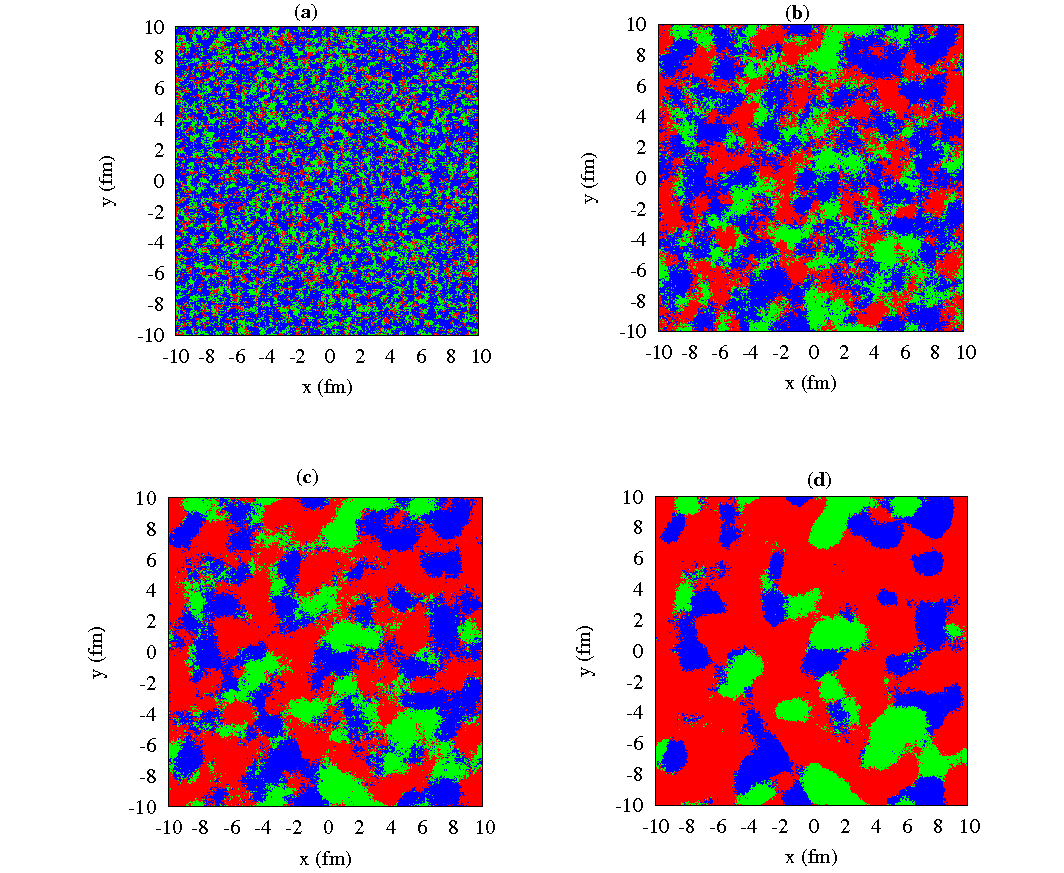}}
\end{center}
\caption{}{(a) Field configurations  at different times with small 
explicit symmetry breaking effects. The shading (color) representing 
dominant region in (d) corresponds to the true vacuum with $\theta = 0$. 
(a) - (d) show the growth of domains for $\tau$ = 1.2, 
1.6, 2.0, and  2.4 fm (with corresponding values of temperature
$T = $ 376, 342, 317, 298 MeV).}
\label{Fig.3}
\end{figure*}

\section{STRONG EXPLICIT SYMMETRY BREAKING AND LARGE FIELD OSCILLATIONS}

 Now we consider the case when explicit symmetry breaking effects make
the initial field configuration completely biased towards $\theta = 0$
direction. Here the initial patch of $l$ rolls down entirely towards
$\theta = 0$ direction with angular variations decreasing during the roll
down. The dynamics of transition is entirely different in this case.
Clearly there is no possibility of different Z(3) domains here, hence
no Z(3) interfaces, or QGP strings will form. We also do not see any
bubble like structures here as were seen in Fig.2. Instead we find
the $l$ settles down to the true vacuum after undergoing huge 
oscillations, with large length and time scales. 

These large oscillations are very similar to coherent oscillations of
the inflaton field in the context of inflation in the early Universe
\cite{infl}. For the inflation, decay of such coherently oscillating field to
particles can be via parametric resonance leading to novel features
in the reheating of the Universe after inflation. Existence of similar
oscillations here raises possibilities of parametric resonance for RHICE
and similar novel dynamics of particle production during such early stages 
of QGP evolution. As for the universe, new possibilities of thermalization 
and symmetry changes may arise here. We hope to explore these issues
in a future work.  

  A direct implication of the existence of such huge oscillations of 
$l$ in the context of RHICE is its possible
effects on the growth of flow anisotropies. In the hydrodynamical
studies of elliptic flow in non-central collisions it is known that 
much of the flow anisotropy develops during early stages of the 
plasma evolution \cite{flow,expn}. In such studies one starts with
equilibrium QGP phase where flow develops due to pressure
gradients. In view of the possibility of large oscillations
during early stages of transition to the QGP phase, 
the equilibrium starting point of these hydrodynamics simulations becomes
suspicious. Our present study does not allow us to address this
issue in the context of hydrodynamical evolution. However,
even with the field theoretical simulation in this work, we can
do a comparative study of momentum anisotropy development
with and without the presence of large oscillations of the
order parameter field $l$. For this we proceed as follows.

  First we need to model the initial QGP system with appropriate
spatial anisotropies. One cannot then use the square lattice with
uniform temperature. We use the temperature profile of Woods-Saxon
shape with the size in the X and Y directions being different
representing elliptical shape for a non-central collision. This allows
us to have a well defined size for the central QGP region, with
temperature smoothly decreasing at the boundary of this region. The 
transverse size $R$ of this system (i.e. profile of temperature) is taken to 
increase with uniform acceleration of 0.015 c per fm, starting from 
an initial value of $R$ equal to the nuclear radius \cite{expn}. The 
initial transverse expansion velocity is taken to be zero. This expanding
background of temperature profile is supposed to represent the
hydrodynamically expanding quark-gluon plasma in which the 
evolution of the order parameter field $l$ will be studied. 
It may appear confusing as $l$ is expected to represent the QGP
phase. Indeed, the normalization of the effective potential in
Eq.(5) from refs.\cite{psrsk,psrsk2} is carried out precisely
so that it represents energy density and pressure of gluons
plus quark degrees of freedom. Still, the dynamics of $l$ from
Eq.(5) does not carry the information of hydrodynamical degrees
of freedom. Various particle modes in $\l$ need to be excited,
which should reach equilibrium, and only then we can expect
some sort of hydrodynamical evolution. Clearly the initial
field configurations assumed here are far from representing
such a hydrodynamical state. A consistent interpretation of
our simulation can be that we are studying long wavelength
modes of $l$ which are couple to a background of short 
wavelength modes which are in thermal equilibrium. This
equilibrated background is expanding with velocity as mentioned
above, and drives the evolution of large wavelength modes of
$l$ via Eq(6).
 
 With this interpretation, our task is straightforward. The
central temperature of the Woods-Saxon profile is taken to
decrease by assuming that the total entropy (integrated in the
transverse plane) decreases linearly as appropriate for Bjorken
dynamics of longitudinal expansion. Note that, with the transverse
expansion being non-zero now, the central temperature will decreases
faster than $\tau^{-1/3}$.  We show, in Fig.4, a sequence
of surface plots of the magnitude of $l$ showing huge oscillations 
with large length scale during quench. Lattice here is again
2000 $\times$ 2000  but we take a large value of $\Delta x = 0.25$ fm
so that the physical lattice size is 50 fm $\times$ 50 fm. The
Woods-Saxon temperature profile (representing QGP region) is taken to 
have a diameter of about 16 fm as appropriate for Au-Au collision for 
RHICE. Large physical size of the lattice allows for the evolution of 
the QGP region to be free from boundary effects. 

\begin{figure*}[!hpt]
\begin{center}
\leavevmode
\epsfysize=8truecm \vbox{\epsfbox{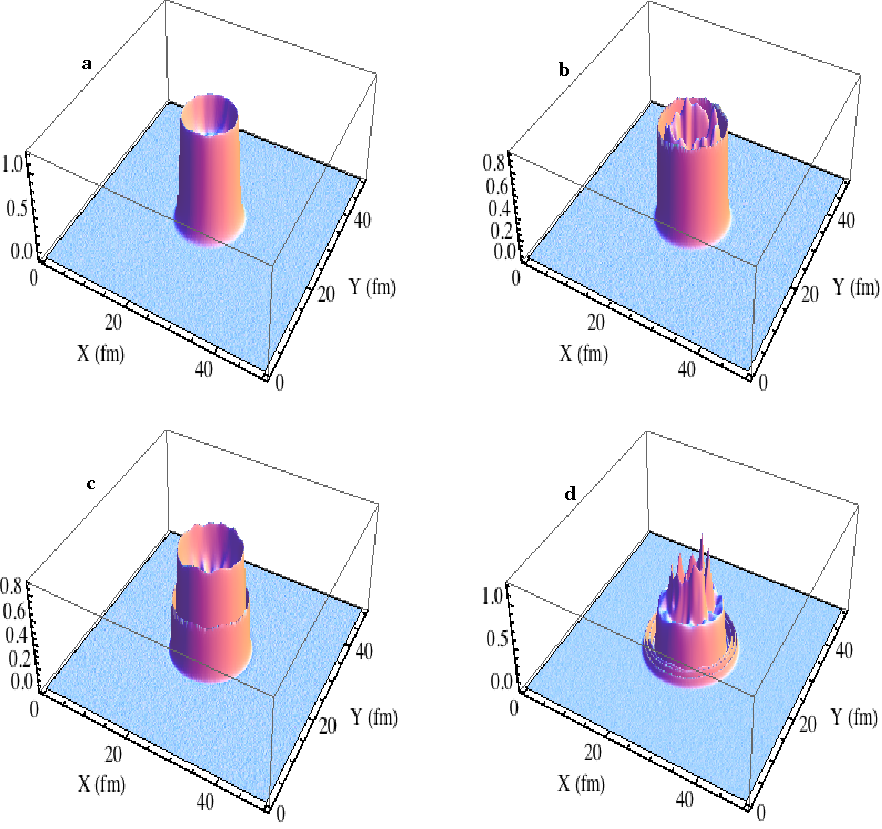}}
\end{center}
\caption{}{Surface plots of the magnitude of $l$ with circular 
geometry for the QGP region. (a) - (d) show 
large oscillations of $l$ during the quench when $l$ rolls down 
everywhere towards the same vacuum with $\theta = 0$. Plots in
(a)-(d) correspond to $\tau = $ 2.6, 3.6, 4.4, 6.2 fm/c, with 
corresponding values of temperature being $T = $ 281, 251, 228, 
191 MeV, respectively}
\label{Fig.4}
\end{figure*}

At any stage we can calculate
the energy momentum tensor $T^{\mu \nu}$ of $l$. We then calculate
the spatial eccentricity $\epsilon_x$  of the $l$ field 
configuration in the standard way,

\begin{equation}
\epsilon_x = {\int dxdy (y^2 - x^2) \rho \over 
\int dxdy (y^2 + x^2) \rho}\\
\end{equation}

 Where $\rho$ is the energy density. We can also calculate the momentum 
density at any time using $T^{0x}$ and $T^{0y}$. Using these components 
we know the momentum density vector at every stage. By integrating it in 
angular bins we calculate its various Fourier coefficients, in particular 
the elliptic flow coefficient $v_2$. 

\section{EFFECTS OF LARGE $l$ OSCILLATIONS ON FLOW ANISOTROPY}

  To study the effects of large $l$ oscillations on flow anisotropy, we 
consider two separate cases. Note that now we are considering the cases with
explicit symmetry breaking, with its effect being strong on the initial
field configuration. First we consider the quench case as described
above. Here, we start with the initial field configuration corresponding
to a small patch near the equilibrium point of $T = 0$ effective 
potential. The patch is taken, as above, with field magnitude randomly
varying in angle between 0 and $2\pi$ with random amplitude uniformly
varying from 0 to  0.01 times the vev of $l$ but now shifted by a constant 
value of 0.011 times the vev along $\theta = 0$ direction. This simulates
the effect of strong explicit symmetry breaking as the entire patch rolls
down towards $\theta = 0$ direction.  This patch is then evolved with the 
temperature profile of Woods-Saxon shape as described above with the 
central temperature having initial value equal to $T_0 = 400$ MeV. As a 
sample case we give results for the case when the initial (elliptical shaped) 
temperature profile  has an eccentricity of 0.5 (with the major and minor
axes being along the x and y axes respectively), which for uniform energy 
density will correspond to $\epsilon_x = -0.143$ from Eq.(7). The initial 
field configuration rolls down in the entire central region towards
roughly $\theta = 0$ direction leading to strong $l$ oscillations.
$\epsilon_x$, and flow coefficients, e.g. $v_2$ are calculated at each 
stage during the evolution. This is shown in Fig.5. In this section we
take the lattice to be 1000 $\times$ 1000 with the physical size of
25 fm $\times$ 25 fm. This still allows for sufficient separation
of the QGP region (of size 16 fm diamater) from the boundary. Here we show 
plots (both for $\epsilon_x$ and for $v_2$)
for two different realizations of the random initial field configuration
(shown as solid and dashed plots). Here, and in all the figures below,
we will show plots for a time upto $\tau = $ 10 fm/c, starting
with $\tau = \tau_0 = 1$ fm/c at the initial stage.
The central temperature decreases (now faster than $\tau^{-1/3}$ due to
nonzero transverse expansion) from an initial 
value $T = T_0 =$ 400 MeV at $\tau = \tau_0 = 1$ fm/c to the final value 
$T = $ 147 MeV at $\tau = $ 10 fm/c. Note from the difference in solid 
and dashed plots in Fig.5 that differences in the initial 
field configuration, which have very small magnitudes, lead to huge 
differences in the values of $\epsilon_x$ and $v_2$.

\begin{figure*}[!hpt]
\begin{center}
\leavevmode
\epsfysize=5truecm \vbox{\epsfbox{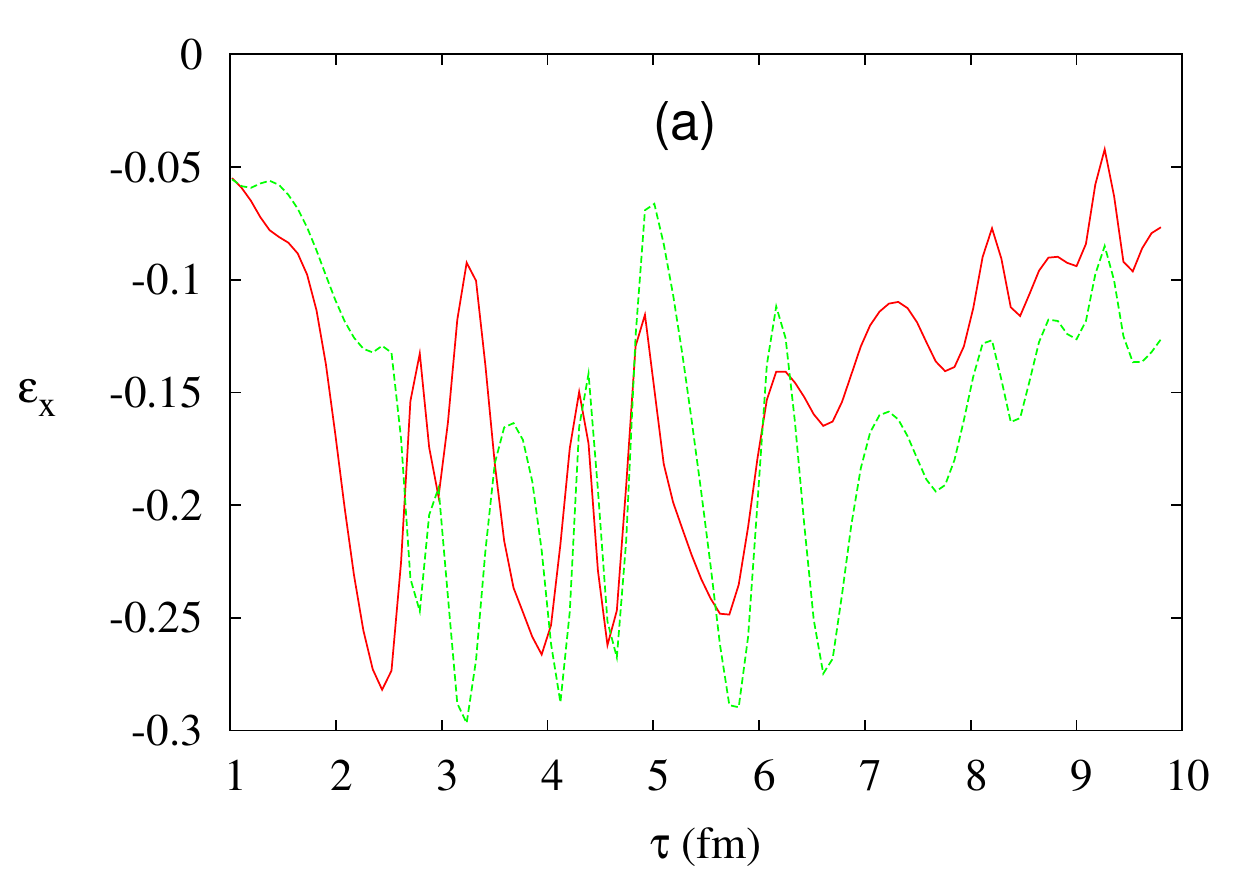}}
\epsfysize=5truecm \vbox{\epsfbox{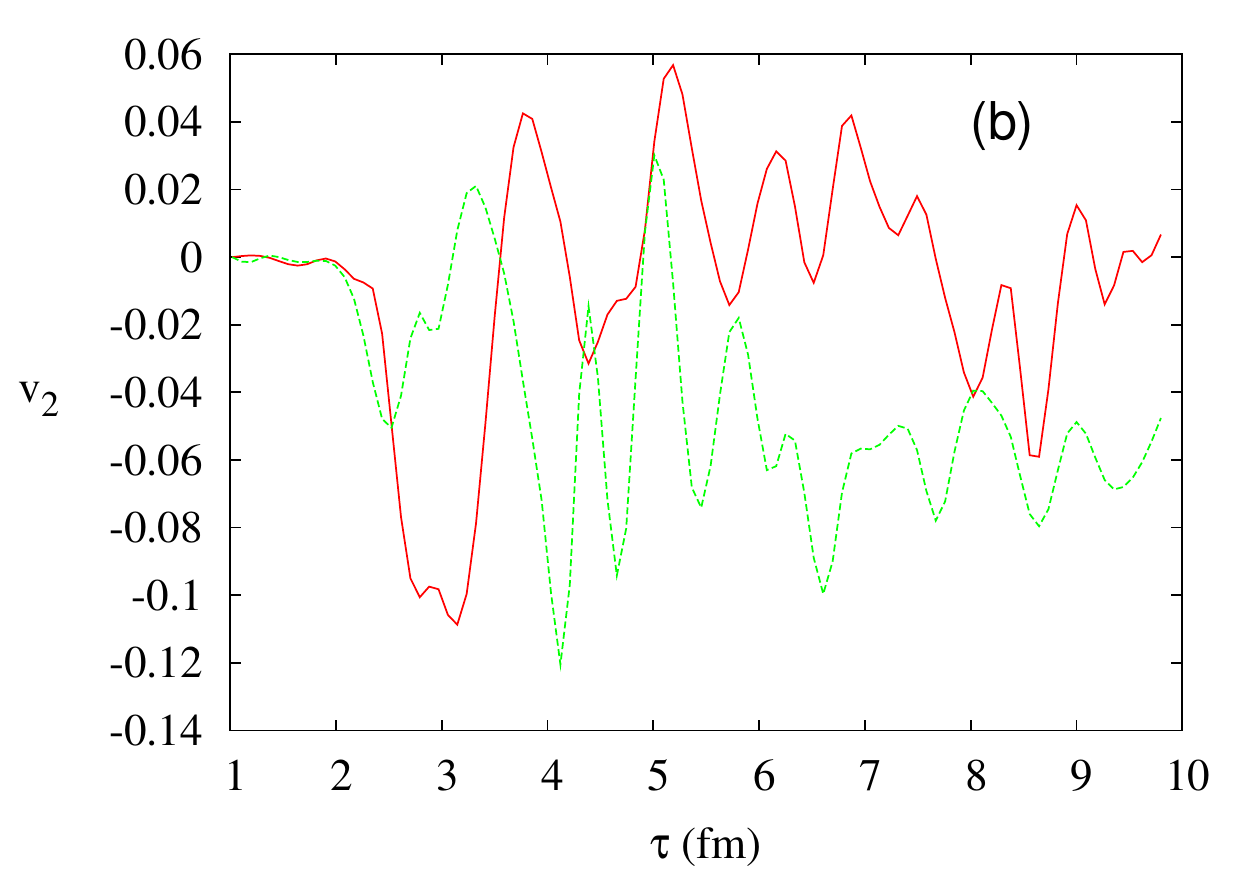}}
\end{center}
\caption{}{These plots correspond to quench case when initial elliptical 
shaped temperature profile has eccentricity of 0.5. Solid and dashed plots
correspond to two different realizations of the initial random
field configuration. (a) and (b) show plots for $\epsilon_x$
and elliptic flow $v_2$ respectively.}
\label{Fig.5}
\end{figure*}

  This situation should be compared to the case of equilibrium
initial conditions, as appropriate for conventional hydrodynamical
simulations. This equilibrium initial condition is implemented
in the following manner here. For the initial temperature profile
of Woods-Saxon shape (again, with a given eccentricity), 
with initial temperature $T = T_0 = 400$ MeV,
we first need to determine the appropriate $l$ configuration which
assumes vacuum expectation value everywhere depending on the
local value of the temperature. To achieve this, we first evolve
field configuration for 1000 time steps with highly dissipative
dynamics while keeping the temperature profile to remain fixed
at the initial profile. A reasonably smooth initial test profile of $l$
is used as the initial field configuration. With highly dissipative
evolution, the field everywhere settles to the local minimum of
the potential quickly. The final configuration is found to be
reasonably independent of the initial configuration assumed
for $l$ as long as it is smooth. This final configuration 
has correct profile as appropriate for the Woods-Saxon profile
of temperature representing lowest energy configuration
everywhere. In order to make a suitable comparison with the
quench case we must incorporate small fluctuations around
this {\it equilibrium} configuration everywhere. For this
purpose we add to the value of the field everywhere, a small 
fluctuating field component with randomly varying angle between 0 
and $2\pi$ and with random amplitude uniformly varying from 0 to  
0.01 times the vev of $l$ (as for the quench case). 
This new field configuration represents the equilibrium
field configuration everywhere, with small fluctuations. This is
taken as the initial field configuration for subsequent
evolution where now any extra dissipation is switched off. This 
stage is taken as representing the initial time $\tau = \tau_0 = $ 1 
fm. The field now evolves with the field equations, Eq.(6). 
Temperature profile also is now allowed to change in time as 
mentioned above.  $\epsilon_x$, and $v_2$ etc. are calculated at
each stage. Fig.6 shows these plots for this equilibrium case,
starting from the time slightly after when extra dissipation (to 
achieve equilibrium configuration) has been switched off.
When random fluctuating component is introduced at every lattice site 
after the end of dissipative evolution, it introduces large gradients.
Thus, for a very short time, there are  large fluctuations as the 
field smoothens to certain level. Thus we show
plots slightly after (by about 0.2 fm/c) the introduction
of random field component.

\begin{figure*}[!hpt]
\begin{center}
\leavevmode
\epsfysize=5truecm \vbox{\epsfbox{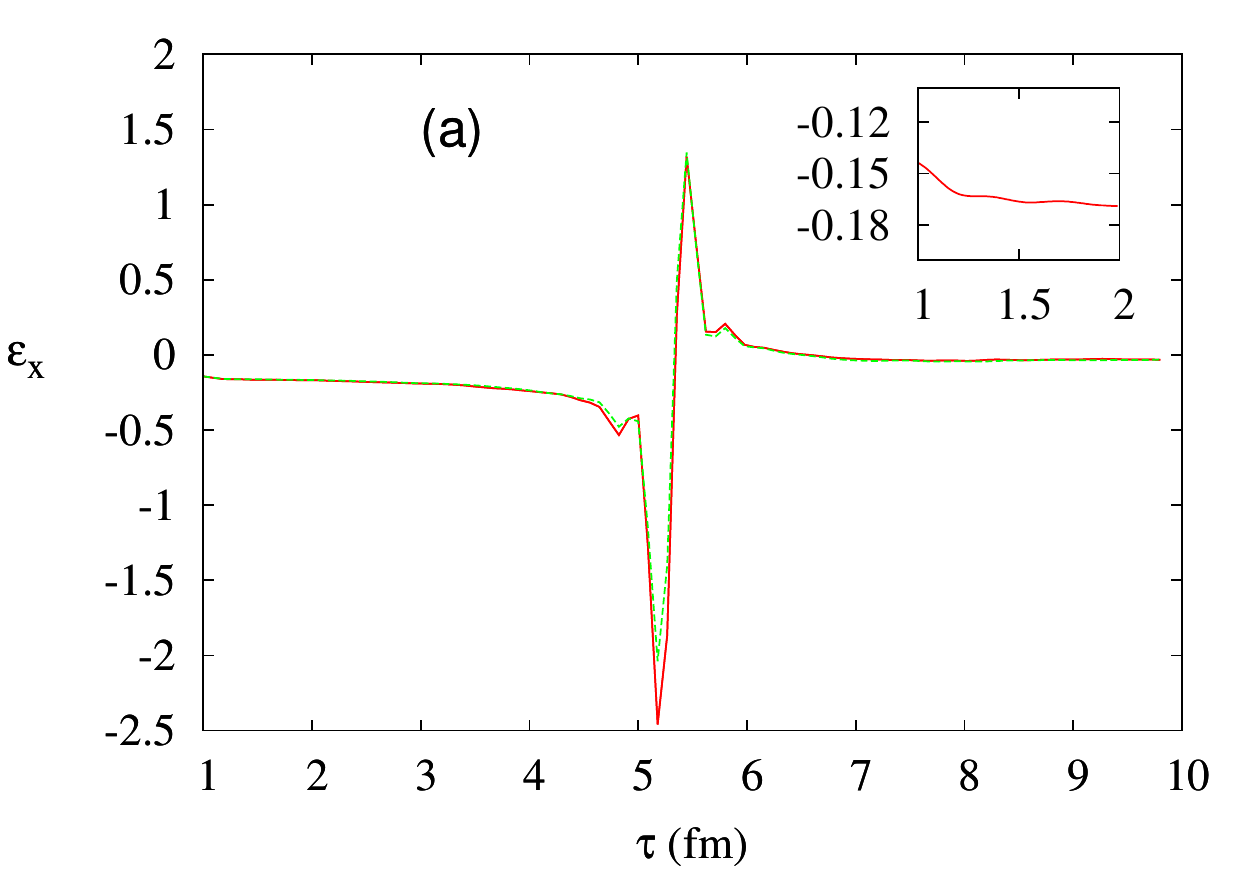}}
\epsfysize=5truecm \vbox{\epsfbox{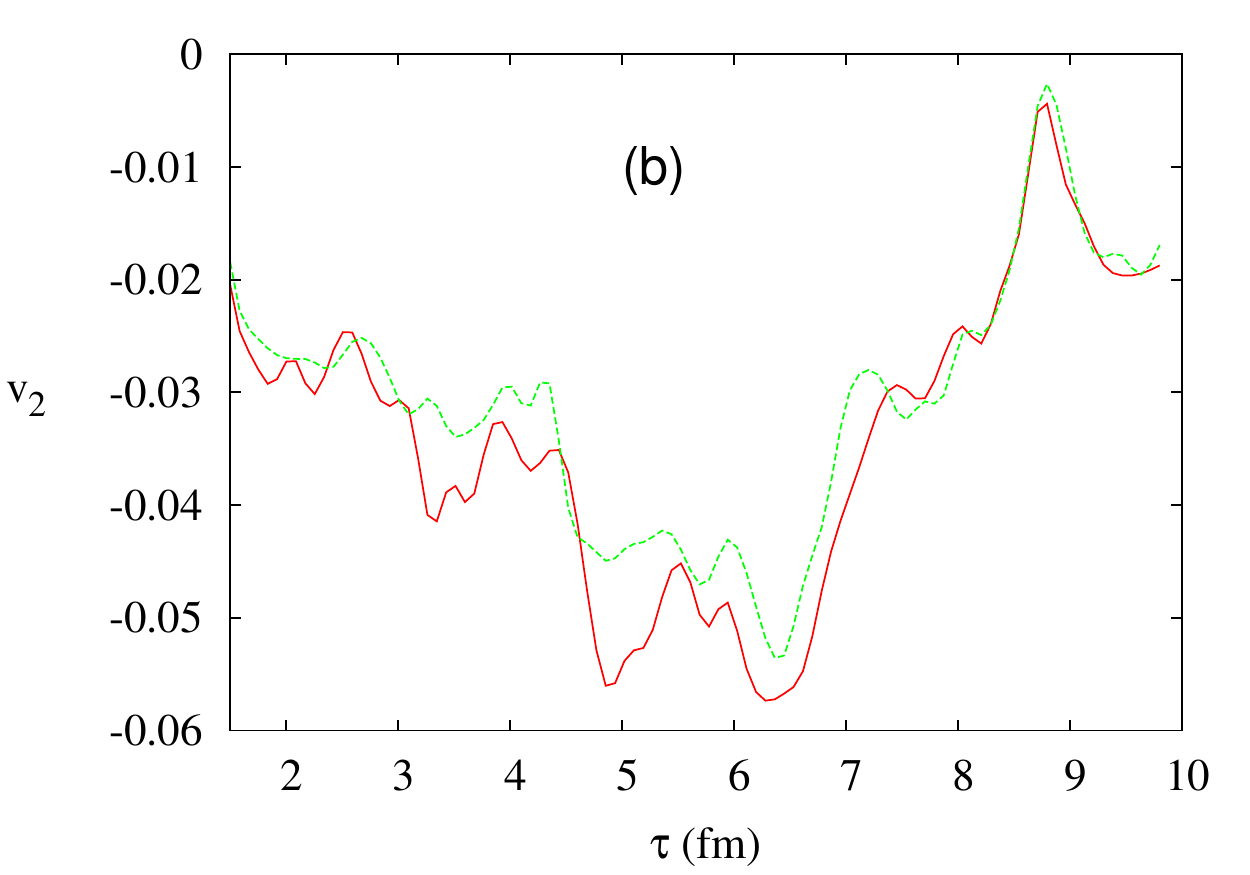}}
\end{center}
\caption{}{Plots as in Fig.5, now for the equilibrium case.
Solid and dashed plots show different realizations of the 
small random fluctuating part of the equilibrium field
configuration. The temperature profile has same initial
eccentricity of 0.5 (with corresponding value of $\epsilon_x
= -0.143$) as in Fig.5. (a) and (b) show plots of $\epsilon_x$ 
and elliptic flow $v_2$ respectively. Note, the initial value
of $\epsilon_x$ for the equilibrated configuration in (a) is
very close to -0.143 as shown by the plot in the inset.} 
\label{Fig.6}
\end{figure*}

 Comparison of Fig.5 and Fig.6 shows the 
dramatic effects of quench induced oscillations on flow anisotropies. 
First note that for the equilibrium case the initial value of 
$\epsilon_x$ is close to the value -0.143 (as shown by the inset in Fig.6)
which exactly corresponds to the initial eccentricity of 0.5 for the 
temperature profile. This gives us confidence that our procedure of 
achieving equilibrated configuration works well.
In contrast, initial value of $\epsilon_x$ for the case of quench
in Fig.5 is very different showing the importance of fluctuations for
this case. We further see huge fluctuations in the values of $\epsilon_x$
and $v_2$ in Fig.5 compared to the equilibrium case shown in Fig.6. 
Note that though there are oscillations in $v_2$ for the equilibrium 
case also, they do not change much from one event to another. In contrast,
for the quench case in Fig.5, there are huge variations between
the two different realizations, i.e. different events. These large 
fluctuations in Fig.5 arise due to large oscillations in $l$ which
itself depends on the nature of randomness in the initial field
configuration. Note that plots for $\epsilon_x$ in Fig.6 show a sharp 
change near $\tau = 5.5$ fm after which $\epsilon_x$ settles down to a 
value close to zero. The central temperature at that
stage is about 208 MeV. Presumably a large part of the region (away from
the central part where temperature is lower due to a Woods-Saxon profile)
may be undergoing transition at that stage leading to large changes in 
field dynamics. We have checked with the contour plots that the 
ellipticity of the profile of $l$, as well as that of energy density, 
does not undergo rapid changes during this stage. (In fact this is
a reflection of a shortcoming in our model where the temperature
profile is taken to have definite initial eccentricity, and its
further expansion is with definite acceleration as discussed above,
starting with zero transverse velocity. A more appropriate 
model may be to take the time evolution of the spatial eccentricity
of the temperature profile from hydrodynamical models and study
the evolution of $l$ using that.) We note strong fluctuations
in both these quantities around this stage, which may be responsible for these
large changes in $\epsilon_x$ at this stage. We hope to develop better
understanding of the dynamics during this stage and discuss in a
followup work. At present what is important to note is that,
apart from this peak region, everywhere else $\epsilon_x$ shows
rather stable values settling down to a value close to zero after $\tau
\simeq$ 5.5 fm, and does not fluctuate much, as compared to
the quench case in Fig.5a.

To further illustrate the importance of fluctuations for the quench 
case, we show plots for the case with
zero eccentricity of the QGP region (i.e. for the temperature 
profile) which also will mean zero value of $\epsilon_x$ for 
uniform density case. Thus, any non-zero $\epsilon_x$
and $v_2$ arise only from the randomness in the initial field configuration.
Fig.7a and Fig.7b show the plots for the equilibrium case. We note
that $\epsilon_x$ and $v_2$ remain very small, apart from a large
change near $\tau = 5.5$ fm for $\epsilon_x$, just as in Fig.6. 
Again, at present what is important to note is that,
apart from this peak region, everywhere else $\epsilon_x$ and $v_2$
remain very small for equilibrium case, as expected for the zero
eccentricity case. Different plots in
Fig.7a and Fig.7b correspond to different realizations of the initial
random field configuration. This situation should be contrasted with
the quenched case as shown in Fig.7c and Fig.7d. Note that $\epsilon_x$
and $v_2$ now fluctuate with large amplitudes, even though eccentricity 
of temperature profile is zero. Further, different realizations of
the initial random field configurations lead to widely different 
plots of these quantities. This shows 
the fluctuating nature of development of flow anisotropies if
large $l$ fluctuations are present. These results suggest
the dynamics of the order parameter field, especially such large
length scale oscillations may play an important role in determining
flow anisotropies and it needs to be incorporated in the hydrodynamical
models.

\begin{figure*}[!hpt]
\begin{center}
\leavevmode
\epsfysize=5truecm \vbox{\epsfbox{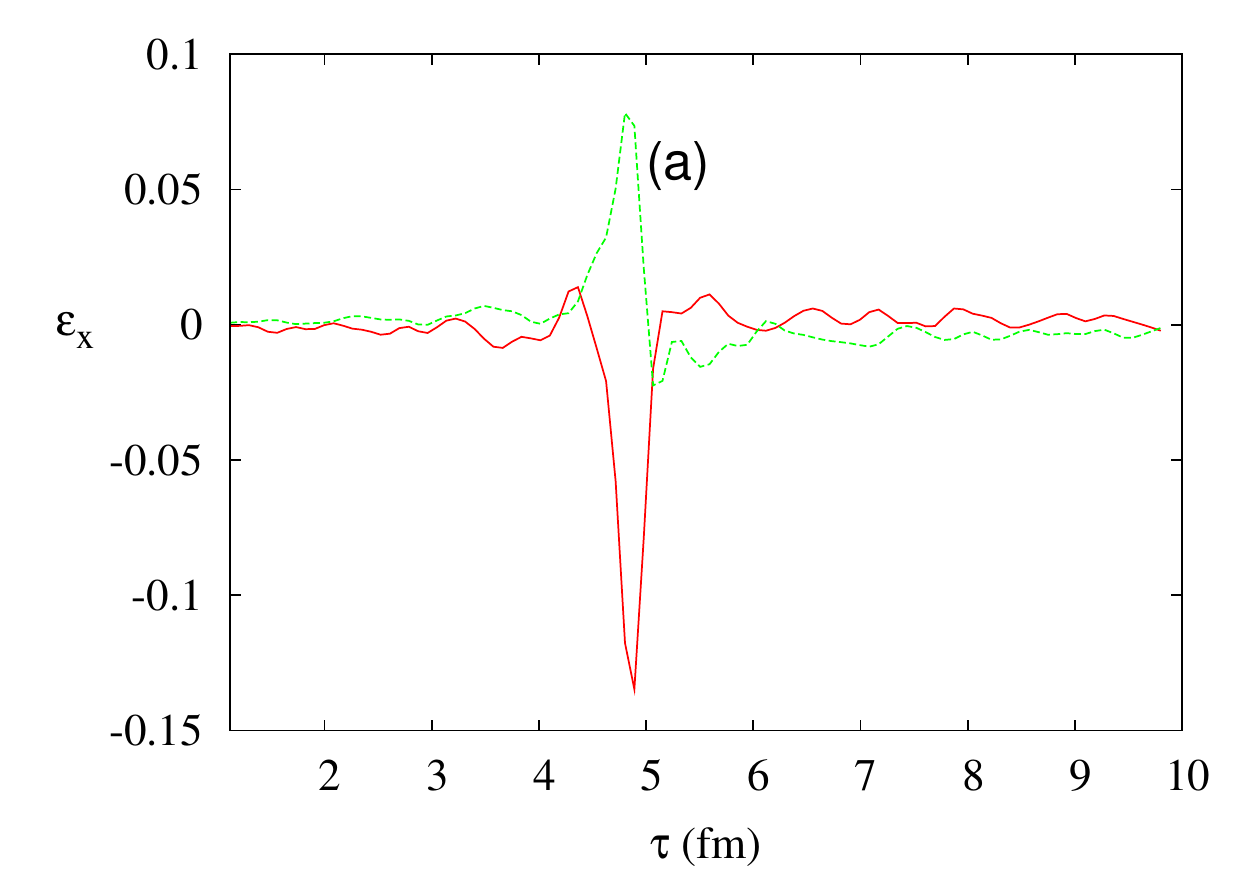}}
\epsfysize=5truecm \vbox{\epsfbox{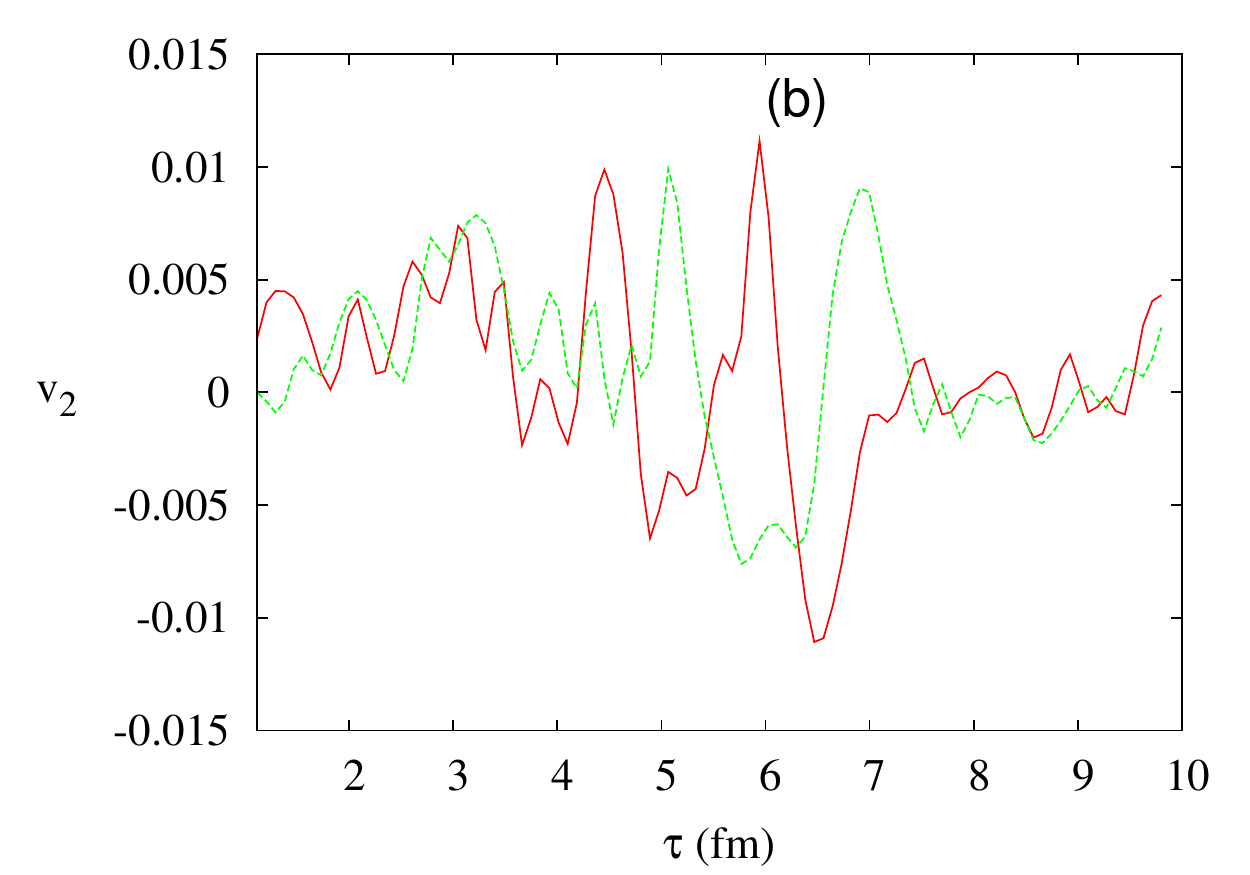}}
\epsfysize=5truecm \vbox{\epsfbox{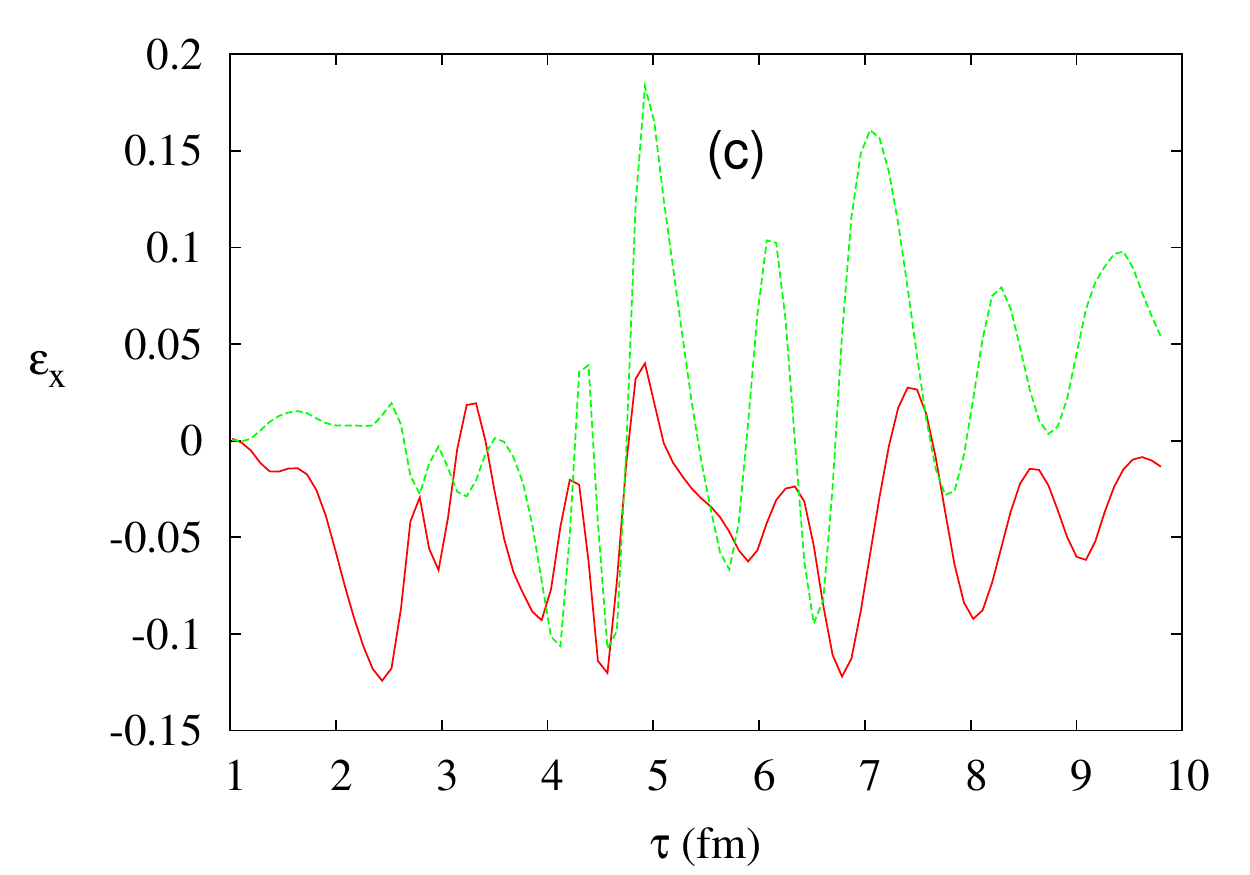}}
\epsfysize=5truecm \vbox{\epsfbox{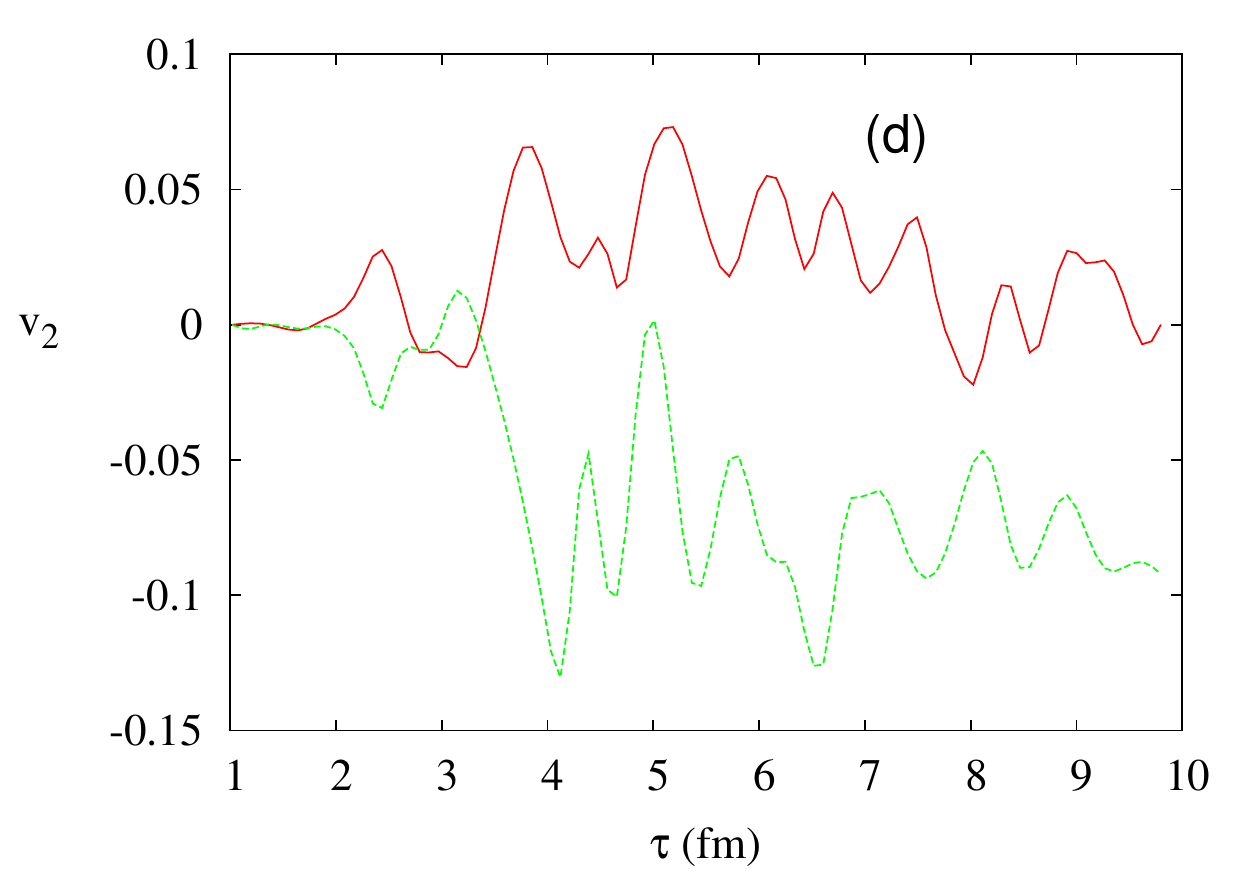}}
\end{center}
\caption{}{These plots are for the zero eccentricity for the 
temperature profile. In all the figures here, solid and dashed plots 
correspond to different realizations of initial random field 
configuration. (a) and (b) show plots of $\epsilon_x$ and $v_2$ for
the equilibrium transition case, while (c) and (d) correspond to
the quench case.}
\label{Fig.7}
\end{figure*}

\section{Conclusions}

 We carry out 2$+$1 dimensional simulation to study the dynamics of 
confinement-deconfinement transition as a quench for the
central rapidity region in relativistic heavy-ion collision 
experiments. We work in the framework of the Polyakov loop model. The 
initial field (in the confining phase) is taken to cover a small 
neighborhood of the confining vacuum $l \simeq 0$ as appropriate for
the initial $T$ = 0 system. This initial field $l$, unable to relax
to the new equilibrium vacuum state with quenched potential at $T = T_0
= 400$ MeV, becomes unstable and rolls down in different directions 
from the top of the central hill in the effective potential of $l$. 
We study the formation of Z(3) domain structure during this evolution. 
When explicit Z(3) symmetry breaking effects (arising from dynamical 
quark effects) are small, then we find well defined Z(3) domains which
coarsen in time. During early stages Z(3) domains (and Z(3) domain walls)
have small sizes of order 1 fm. However at this stage the magnitude of
$l$ is very small, of order few present of its vacuum expectation value.
Domains coarsen rapidly as the magnitude of $l$ grows and by 
the time $\tau = 4-5$ fm, domains are of size several fm.
 Surprisingly, the magnitude plot of $l$ shows vacuum bubble 
like configurations, such as those which arise in a first order transition.
This first order transition like behavior occurs even though there is no 
metastable vacuum separated by a barrier from the true vacuum for the 
parameter values used. These bubble like configurations expand as well, 
somewhat in a similar manner as during a first order transition.  
This result points to new interesting 
possibilities for the phase transition dynamics. For example, such bubble
like structures may lead to a dynamics of phase separation in the
case of the universe similar to the original Witten's scenario 
\cite{witten}, even when there is no underlying first order transition.
This may also have important implications for RHICE. 
This new possibility of transition dynamics needs to be understood
and analyzed in detail, see \cite{bubble} for a study of these issues.  

When the initial patch of $l$ is only partially symmetric around $l = 0$ 
(as appropriate for small explicit symmetry breaking from quark effects), 
the dynamics retains these qualitative aspects, with true vacuum domains 
(with $\theta = 0$) growing dominantly at the cost of the other two 
metastable Z(3) domains. In this case Z(3) walls are fewer and 
relatively smaller, and they disappear more quickly.
When the initial patch  of $l$ (around equilibrium point for $T = 0$ 
effective potential) is significantly shifted towards the true vacuum 
for the quenched $T = T_0$ effective potential (as will happen when
explicit symmetry breaking is strong), then $l$ rolls down roughly 
along the same direction with angular variations becoming
smaller during the roll down. In this case only $\theta = 0$ vacuum
survives and no other Z(3) domains are formed. Also, in this case we 
do not find bubble-like configurations. However, in this case we find 
huge oscillations of $l$ with large length scales. This is similar
to the scenarios of reheating via parametric
resonance in the case of inflation in the early Universe. In our case of
RHICE also it raises important questions about the possibility of novel
modes of particle production from these large oscillations of $l$ during
the early stages of the transition. We have shown that these large $l$
oscillations can strongly affect the evolution of flow
anisotropies in RHICE. The spatial eccentricity and the flow
coefficients, e.g. the elliptic flow $v_2$ are found to undergo large
fluctuations during the evolution of the system. These results suggest
the dynamics of the order parameter field, especially such large
length scale oscillations may play an important role in determining
flow anisotropies and it needs to be incorporated in the hydrodynamical
models.

\acknowledgments

  We are extremely grateful to Umashankar Gupta for his help in
the simulations, especially for the part involving novel bubble
like configurations. We are also very grateful to Rajeev Bhalerao,
Vivek Tiwari, Sanatan Digal, P.S. Saumia, Abhishek Atreya, and
Partha Bagchi for very useful comments and suggestions.

%%%%%%%%%%%%%%%%%%% 

\end{document}